# Optical two-way time and frequency transfer over free space

**Fabrizio R. Giorgetta**[1*], **William C. Swann**[1], **Laura C. Sinclair**[1],
**Esther Baumann**[1], **Ian Coddington**[1] **and Nathan R. Newbury**[1*]

[1]*National Institute of Standards and Technology, 325 Broadway, Boulder, Colorado 80305*
[*]*e-mail: fabrizio@nist.gov; nnewbury@boulder.nist.gov*

The transfer of high-quality time-frequency signals between remote locations underpins a broad range of applications including precision navigation and timing, the new field of clock-based geodesy, long-baseline interferometry, coherent radar arrays, tests of general relativity and fundamental constants, and the future redefinition of the second[1-7]. However, present microwave-based time-frequency transfer[8-10] is inadequate for state-of-the-art optical clocks and oscillators[1,11-15] that have femtosecond-level timing jitter and accuracies below $10^{-17}$; as such, commensurate *optically-based* transfer methods are needed. While fiber-based optical links have proven suitable[16,17], they are limited to comparisons between fixed sites connected by a specialized bidirectional fiber link. With the exception of tests of the fundamental constants, most applications instead require more flexible connections between remote and possibly portable optical clocks and oscillators. Here we demonstrate optical time-frequency transfer over free-space via a two-way exchange between coherent frequency combs, each phase-locked to the local optical clock or oscillator. We achieve femtosecond-scale timing deviation, a residual instability below $10^{-18}$ at 1000 s and systematic offsets below $4\times10^{-19}$, despite frequent signal fading due to atmospheric turbulence or obstructions across the 2-km link. This free-space transfer would already enable terrestrial links to support clock-based geodesy. If combined with satellite-based free-space optical communications, it provides a path toward global-scale geodesy, high-accuracy time-frequency distribution, satellite-based relativity experiments, and "optical GPS" for precision navigation.

In optical transfer over fiber links, convincingly demonstrated over 920 km across Germany[16], the relative clock frequencies between two remote sites are compared continuously via an optical carrier. This same approach could be applied over free-space as well[18,19] but this frequency comparison requires a continuously operating link, which is incompatible with free-space transmission as turbulence will cause significant and frequent signal fading through beam scintillation, beam wander, and angle-of-arrival jitter[20]. Moreover, simple physical obstructions and platform motion will routinely interrupt a free-space link. Therefore, we pursue the optical analog to conventional microwave-based two-way time-frequency transfer (TWTFT). Optical TWTFT compares the elapsed time intervals between two sites rather than comparing their frequencies. Consequently, instead of requiring a continuous link, it only requires an exchange of pulses to synchronize the sites at the start and stop times of the time interval; continuous observation between these start and stop times is unnecessary, although quasi-continuous measurements can decrease the short-term uncertainty.

The measured time interval difference can be used in different ways. For geodesy, two high-accuracy clocks located at different geographical locations will have a difference in time intervals that reflects their relative gravitational redshifts with a sensitivity of $10^{-16}$ per meter[1]. Clock-based geodesy is an intriguing alternative to current time-consuming and labor-intensive leveling for geodetic calibration data[2]. Alternatively, the time-interval difference can be used for jam-free synchronization of remote clocks, for synchronization of oscillators across a coherent sensor array (*e.g.* a long-baseline interferometer or coherent radar array), or to provide position information as in conventional GPS but with much higher precision.

As required to support state-of-the-art optical clocks and oscillators, our optical TWTFT achieves residual timing deviations of 1 femtosecond, and residual uncertainties and bias below $10^{-18}$. Because conventional photodetection of pulse arrival times would lead to picosecond level timing jitter or worse, we implement linear optical sampling (LOS) between pulse trains of optical frequency combs[21] to obtain the femtosecond-level timing. (Balanced optical cross-correlation is a possible alternative[22,23], although it requires higher transmitted powers.) Furthermore, because the optical path-length between sites will vary due to atmospheric turbulence, platform motion and slow changes in air temperature and pressure (with corresponding 1 ppm/°C and 3 ppb/Pa optical path-length changes at sea level), we implement a two-way exchange of pulses from frequency combs located at each site in analogy with rf-based TWTFT. This optical TWTF is highly effective because of the fundamental reciprocity of free-space single-mode links[24,25]. With this approach, optical path-length variations are cancelled below 300-nm over time scales from milliseconds to hours. Moreover, we show that this performance is not limited by variations across the 2-km link distance. As such, similar performance is expected over much longer distances. Finally, as an added benefit, the approach is compatible with free-space optical communications and can therefore exploit existing and forthcoming telecommunication technologies.

The optical TWTFT was demonstrated over a link consisting of 2-km of free-space and 400-m of optical fiber, as shown in Fig. 1. Transmission of comb pulses over optical fiber does preserve the comb timing other than the effects of path-length variations[26], and the use of fiber optics allows the main system to be located in a laboratory far from the free-space launching optics. To quantify the residual transfer noise, the two end sites "A" and "B" are adjacent and use a common optical oscillator. In all other regards, the two sites are independent. At each site, a coherent pulse train (comb) is generated by phase locking a femtosecond fiber laser to the optical oscillator but with the repetition rates differing by $\Delta f_r = 1/\tau$ between sites. As with other dual-comb applications[21], LOS between these mismatched optical pulse trains gives time/phase information of the underlying optical waveform over broad optical bandwidths (1 THz) as a train of interferograms (cross-correlations), nominally separated by $\tau$. The time intervals between the $0^{th}$ and $n^{th}$ interferogram measured at site A and B are

$$T_A(n\tau) = n\tau + \Delta T_{Path}(n\tau) + \Delta T_{AB}(n\tau) \text{ and}$$

$$T_B(n\tau) = n\tau + \Delta T_{Path}(n\tau) - \Delta T_{AB}(n\tau),$$

Where $\Delta T_{Path}$ is the cumulative change in the time of flight over the reciprocal path between the sites, and $\Delta T_{AB}$, the quantity of interest, is the cumulative timing difference between the clock/oscillator at site A and site B. Both $\Delta T_{Path}$ and $\Delta T_{AB}$ are assumed to vary slowly over $\tau$. The two-way time difference is $\Delta T_{AB} = (T_A - T_B)/2$. Note that measurements at only $n = 0$ and at $n = N$ are sufficient to find $\Delta T_{AB}(N\tau)$, provided the combs remain continuously phase locked to the optical oscillator.

The interpretation of $\Delta T_{AB}(n\tau)$ is straightforward; a linear fit yields a slope that is exactly the fractional offset in time or in frequency, given by $\Delta f_{Clock}/f_{Clock}$, where $\Delta f_{Clock}$ is the frequency offset between two clocks with nominal frequency $f_{Clock}$. Here we use a common "clock" so that the fractional offset should be zero in the absence of a systematic bias. The Allan deviation of $\Delta T_{AB}(n\tau)$ expresses the uncertainty on the measured fractional offset.

Figure 2 shows data from one day. The effects of turbulence are evident. Based on the measured angle-of-arrival jitter, the turbulence structure constant is $C_n^2 \sim 10^{-15}$ to $10^{-14}$ m$^{-2/3}$ (Ref.

20). This angular jitter is mainly cancelled through feedback to a steering mirror but the link is still interrupted because of turbulence-induced scintillation, physical obstructions, or loss of steering-mirror tracking. For example, from 12:00 to 13:00 interruptions occurred 23% of the time with durations up to 10 s, while from 16:30 to 17:30, they occurred 0.3% of the time with durations up to 0.5 s. (In our current processing, an interruption corresponds to the received signal fading below 20 nW at the photodiode, or ~1500 photons per pulse.) Nevertheless, both $\Delta T_A \equiv T_A - n\tau$, $\Delta T_B \equiv T_B - n\tau$, and therefore $\Delta T_{AB}$ are tracked through interruptions since the combs remain phase-locked to the underlying clock. However, a slip of the phase lock of a comb does reset the measurement, as at 13:35, which limited measurement periods to a few hours. A zero dead-time counter, more robust phase locks or possibly real-time processing[27] could suppress this problem.

Figure 3 shows timing-jitter power spectral densities (PSD) $S_A(f)$ of $\Delta T_A$ and $S_{AB}(f)$ of $\Delta T_{AB}$. The different contributions to the one-way noise, $S_A(f)$, can be deconstructed. Above 100 Hz, the noise floor is attributed to detector noise, shot noise, and phase jitter on the combs' phase lock to the common clock. Between 20 Hz and 100 Hz, $S_A(f)$ is dominated by vibration-induced fiber noise. (At shorter fiber lengths, building vibrations induce similar noise through motion of the free-space launching optics[28].) Below 20 Hz, the turbulence-induced atmospheric piston effect dominates with a contribution $S_T(f) = 0.016 c^{-2} C_n^2 L V^{5/3} f^{-8/3}$ s$^2$/Hz over *V/L$_0$<f<0.3V/D*, where $L \sim 2$ km, $V \sim 1$ m/s is the wind speed, $c$ is the speed of light, $L_0 \sim 10\text{-}100$ m is the outer scale, and $D \sim 5$ cm is the aperture diameter[20,29]. In stark contrast to $S_A(f)$, the PSD of the residual timing difference, $S_{AB}(f)$, follows the transceiver noise, which is $10^3$ to $10^4$ times lower than $S_A(f)$.

As shown in Fig. 4, the modified Allan deviation for $\Delta T_{AB}$ falls below $10^{-18}$ around a measurement interval of 1,000 seconds and equals the Allan deviation for a shorted link, indicating that transceiver noise, and not residual free-space path-length variations, limits performance. Even if the measurement is restricted to two 100-ms intervals at the start and stop of the overall interval, the resulting instability (equivalent to the overlapping Allan deviation for 100-ms smoothed data) still reaches $10^{-18}$ at ~1,000 seconds. The corresponding timing deviation, shown in Fig. 4b, averages down with the square root of time below 1 second, after which $1/f$ or $1/f^2$ phase noise dominates, presumably from non-common fiber-optic paths in the transceiver.

The instability (Allan deviation) is one measure of performance; it is equally important that the transfer introduces no frequency offsets, *i.e.* non-zero slope to $\Delta T_{AB}$. As shown in Fig. 4c, the bias, or fractional offset, is below $\sim 4 \times 10^{-19}$ (one-sigma) for the two-way measurement. In contrast, one-way transfer, $\Delta T_A$, can exhibit a significant offset. For example, a linear 10 °C temperature increase per hour over 2 km results in a $1.4 \times 10^{-14}$ offset, much larger than the Allan deviation; platform motion would lead to even higher offsets through Doppler shifts.

The current demonstration covers 2 km but as shown above is not distance-limited, and equivalent performance out to much longer terrestrial links should be possible. Very long distance or global coverage will require ground-to-satellite or satellite-to-satellite optical TWTFT, and could replace conventional or future microwave-based approaches[8-10] for high-performance applications. While the integrated turbulence from ground to satellite is similar to that encountered across this 2-km link, clearly additional issues arise in satellite-based optical TWTFT: longer delays tax the reciprocity condition; large Doppler shifts and platform displacements must be handled; and a coherent link must be maintained, even if intermittently. Fortunately, satellite-based optical communications is an active and fruitful area of development[20,25,30]. The optical TWTFT demonstrated here is compatible with these advancements, and a marriage of the two could usher in new possibilities in ultra-precise and accurate global time/frequency distribution, global geodesy, and satellite-based experiments on general relativity.

This work was funded by the DARPA DSO QuASAR program and by the National Institute of Standards and Technology (NIST). We acknowledge helpful discussions with Scott Diddams, Jean-Daniel Deschênes, Sumanth Kaushik, Steven Michael, Ron Parenti, Tom Parker, Frank Quinlan and Till Rosenband, and assistance from Emily Williams and Alex Zolot.

## References

1. Chou, C. W., Hume, D. B., Rosenband, T. & Wineland, D. J. Optical clocks and relativity. *Science* **329**, 1630–1633 (2010).

2. Bondarescu, R. *et al.* Geophysical applicability of atomic clocks: direct continental geoid mapping. *Geophys. J. Int.* **191**, 78–82 (2012).

3. Müller, J., Soffel, M. & Klioner, S. Geodesy and relativity. *J. Geod.* **82**, 133–145 (2008).

4. Schiller, S. *et al.* Einstein gravity explorer-a medium-class fundamental physics mission. *Exp. Astron.* **23**, 573–610 (2009).

5. Wolf, P. *et al.* Quantum physics exploring gravity in the outer solar system: the SAGAS project. *Exp. Astron.* **23**, 651–687 (2009).

6. Uzan, J. The fundamental constants and their variation: observational and theoretical status. *Rev. Mod. Phys.* **75**, 403–455 (2003).

7. Gill, P. When should we change the definition of the second? *Phil. Trans. R. Soc. A* **369**, 4109–4130 (2011).

8. Bauch, A. *et al.* Comparison between frequency standards in Europe and the USA at the $10^{-15}$ uncertainty level. *Metrologia* **43**, 109–120 (2006).

9. Samain, E. *et al.* The T2L2 ground experiment time transfer in the picosecond range over a few kilometres. In *Proceedings of the 20th European Frequency and Time Forum*, pages 538–544 (2006).

10. Cacciapuoti, L. & Salomon, C. Space clocks and fundamental tests: The ACES experiment. *Eur. Phys. J. Special Topics* **172**, 57–68 (2009).

11. Chou, C. W., Hume, D. B., Koelemeij, J. C. J., Wineland, D. J. & Rosenband, T. Frequency comparison of two high-accuracy $Al^+$ optical clocks. *Phys. Rev. Lett.* **104**, 070802 (2010).

12. Katori, H. Optical lattice clocks and quantum metrology. *Nature Photon.* **5**, 203–210 (2011).

13. Ludlow, A. D. *et al.* Sr lattice clock at $1\times10^{-16}$ fractional uncertainty by remote optical evaluation with a Ca clock. *Science* **319**, 1805–1808 (2008).

14. Fortier, T. M. *et al.* Generation of ultrastable microwaves via optical frequency division. *Nature Photon.* **5**, 425–429 (2011).

15. Jiang, Y. Y. *et al.* Making optical atomic clocks more stable with $10^{-16}$-level laser stabilization. *Nature Photon.* **5**, 158–161 (2011).

16. Predehl, K. *et al.* A 920-kilometer optical fiber link for frequency metrology at the 19th decimal place. *Science* **336**, 441–444 (2012).

17. Lopez, O. *et al.* Simultaneous remote transfer of accurate timing and optical frequency over a public fiber network. *http://arxiv.org/ftp/arxiv/papers/1209/1209.4715.pdf* pages 1–4 (2012).

18. Sprenger, B., Zhang, J., Lu, Z. H. & Wang, L. J. Atmospheric transfer of optical and radio frequency clock signals. *Opt. Lett.* **34**, 965–967 (2009).

19. Djerroud, K. *et al.* Coherent optical link through the turbulent atmosphere. *Opt. Lett.* **35**, 1479–1481 (2010).

20. Andrews, L. C. & Phillips, R. L. *Laser Beam Propagation through Random Media* (SPIE Press, 1998).

21. Coddington, I., Swann, W. C., Nenadovic, L. & Newbury, N. R. Rapid and precise absolute distance measurements at long range. *Nature Photon.* **3**, 351–356 (2009).

22. Kim, J., Cox, J. A., Chen, J. & Kärtner, F. X. Drift-free femtosecond timing synchronization of remote optical and microwave sources. *Nature Photon.* **2**, 733–736 (2008).

23. Lee, J., Kim, Y. J., Lee, K., Lee, S. & Kim, S. W. Time-of-flight measurement with femtosecond light pulses. *Nature Photon.* **4**, 716–720 (2010).

24. Shapiro, J. H. Reciprocity of the turbulent atmosphere. *J. Opt. Soc. Am.* **61**, 492–495 (1971).

25. Parenti, R. R., Michael, S., Roth, J. M. & Yarnall, T. M. Comparisons of $C_n^2$ measurements and power-in-fiber data from two long-path free-space optical communication experiments. *Proc. of SPIE* **7814**, 78140Z (2010).

26. Marra, G. *et al.* High-resolution microwave frequency transfer over an 86-km-long optical fiber network using a mode-locked laser. *Opt. Lett.* **36**, 511–513 (2011).

27. Roy, J., Deschênes, J.-D., Potvin, S. & Genest, J. Continuous real-time correction and averaging for frequency comb interferometry. *Opt. Express* **20**, 21932–21939 (2012).

28. Giorgetta, F. R. *et al.* Two-way link for time interval comparison of optical clocks over free-space. In *CLEO: Applications and Technology. postdeadline*, page CTh5D.10 (Optical Society of America, 2012).

29. Conan, J., Rousset, G. & Madec, P.-Y. Wave-front temporal spectra in high-resolution imaging through turbulence. *J. Opt. Soc. Am. A* **12**, 1559–1570 (1995).

30. Koishi, Y. *et al.* Research and development of 40Gbps optical free space communication from satellite/airplane. In *International Conference on Space Optical Systems and Applications (ICSOS)*, pages 88 –92 (2011).

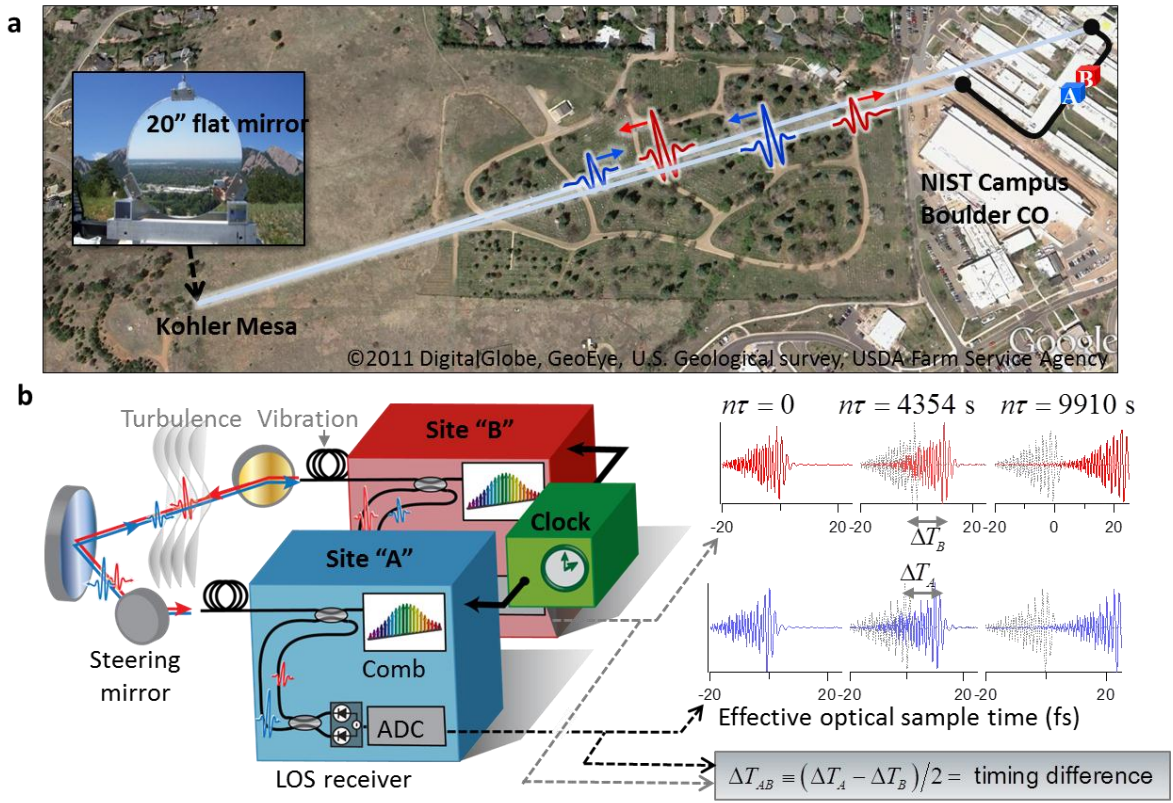

Figure 1: Setup for optical two-way frequency/time transfer (TWTFT). **a,** The transfer takes place between two co-located sites "A" and "B" sharing a common optical oscillator or "clock", allowing evaluation of the residual timing deviation. The sites are linked by two 200-m optical fiber paths from the laboratory to the free-space launch points (black line) and a 2-km air path (grey line). **b,** A frequency comb, consisting of a femtosecond fiber laser phase locked to the clock, generates a pulse train that is coherent, i.e. "ticks" synchronously, with the clock at a repetition rate $f_r$ ~100 MHz for Site A and $f_r+\Delta f_r$ ~100.003 MHz for Site B. A portion of the pulse train is transmitted while the remainder optically samples the received pulse train to generate an interferogram (red, blue traces) at an effective time step of ~$\Delta f_r/f_r^2$ per pulse. Spectral processing yields the time difference, $\Delta T_A$ or $\Delta T_B$, on the $n^{th}$ interferogram from its "expected"

unperturbed arrival time (grey traces) at sites A and B. The timing difference between the sites, $\Delta T_{AB}=(\Delta T_A-\Delta T_B)/2$, is independent of optical path-length variations.

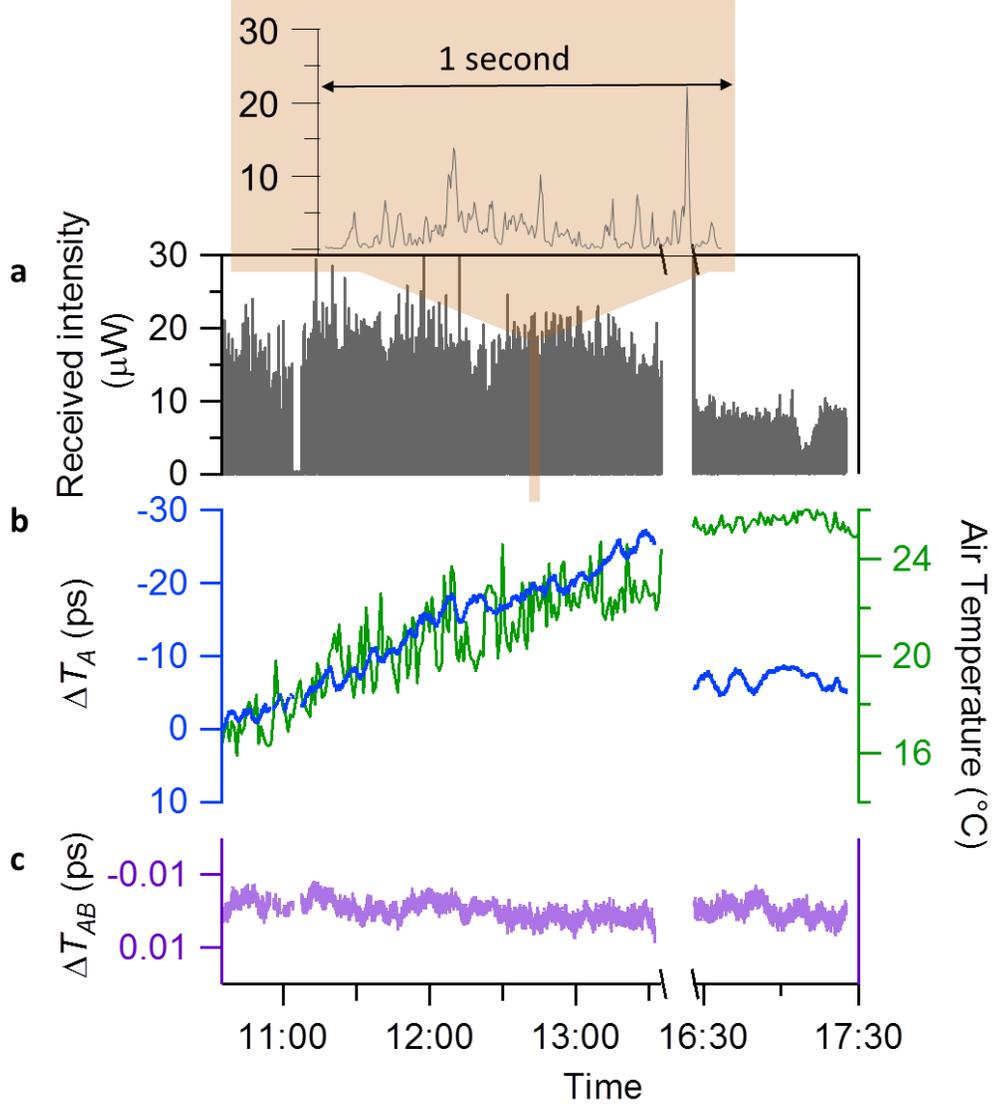

Figure 2: Example data. **a**, The detected intensity at site A exhibits the strong fluctuations and fading characteristic of coherent links over turbulent paths. **b**, Variation in the one-way time-of-flight, $\Delta T_A$, (blue, left axis) and air temperature (green, right axis). The reset in $\Delta T_A$ at 13:45 is a consequence of a phase slip between the clock and one comb. **c**, Residual timing difference, $\Delta T_{AB}$, acquired at $\tau = 300$ μs intervals and averaged over 100-ms time intervals. The slow ripple (standard deviation of 2.5 fs) is attributed to temperature-induced path-length fluctuations within the transceiver and not the common-mode free-space link.

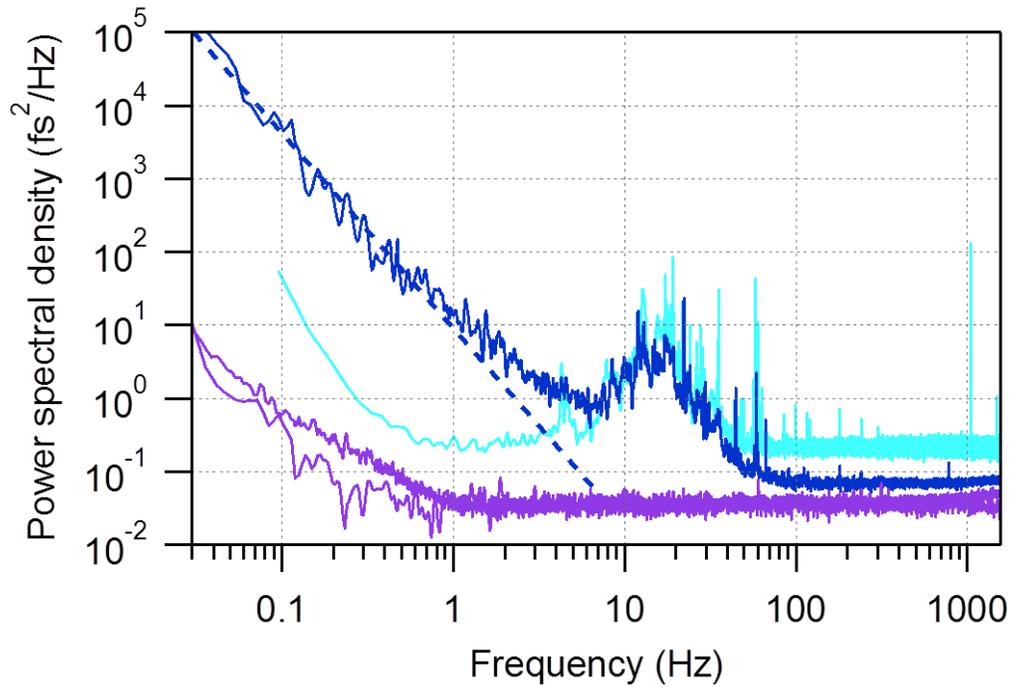

Figure 3: Power spectral densities (PSDs). The PSD for the time-of-flight, $\Delta T_A$, across 2 km of air and 400 m of optical fiber (dark blue line) has contributions from the 400-m optical fiber (light blue line) at intermediate frequencies and the atmospheric piston effect (dashed blue line) at low frequencies, calculated for 1-m/s wind speed and $C_n^2 = 2.5 \times 10^{-14}$ m$^{-2/3}$. In contrast, the PSD for the two-way timing difference, $\Delta T_{AB}$, (dark purple) has negligible contribution from atmospheric turbulence or fiber noise and lies directly on top of the two-way PSD measured over a shorted path (light purple line).

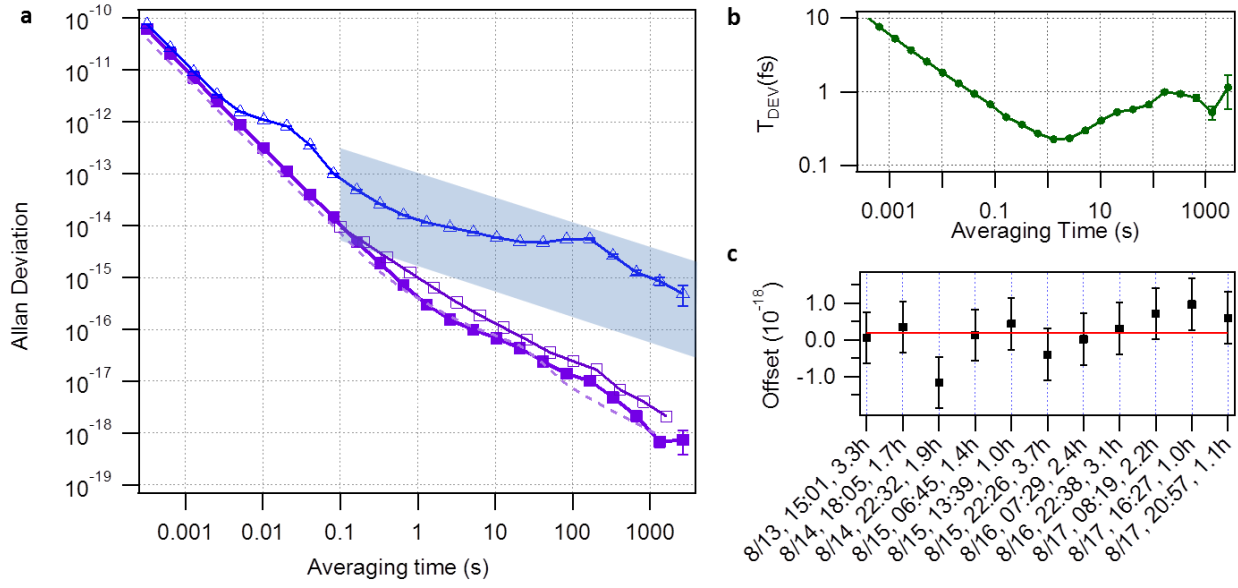

Figure 4: Precision (Allan deviation) and offset of the optical TWTFT, evaluated over multiple data sets covering a total of 24 hours acquisition. **a**, The modified Allan deviation for the optical TWTFT, $\Delta T_{AB}$ (solid squares) is well below the instabilities of the state-of-the-art optical clocks (shaded region)[1,11-16] and lies directly on the modified Allan deviation for a shorted link (dashed line). The overlapping Allan deviation at 100 ms averaging (open squares) corresponds to two isolated 100-ms measurements at the start and end of the time interval and exhibits similar performance. In contrast, the modified Allan deviation for $\Delta T_A$ (open triangles) is significantly higher and limited by optical path length fluctuations. **b**, The timing deviation of the two-way measurement, $\Delta T_{AB}$. **c**, The fractional offset, or bias, across eleven data sets (squares) has a weighted average (solid line) of $1.8 \pm 2.1 \times 10^{-19}$, consistent with zero and well below the best optical clocks. (The uncertainty per point is conservatively estimated at $7 \times 10^{-19}$ assuming a flat Allan deviation at integration times beyond 2000 s.)